\documentclass{aastex63}

\def\bfc{}

\def\kms{km~s$^{-1}$~}
\def\cm3{cm$^{-3}$~}

\usepackage{graphicx}
\usepackage{epsf}
\usepackage{natbib}
\usepackage{enumitem}
\usepackage{amsmath}

\newcommand{\simlt}{\lower.5ex\hbox{$\; \buildrel < \over \sim \;$}}

\providecommand{\sorthelp}[1]{}

\begin{document}

\title{Extraplanar gas in Edge-on Galaxies traced by SOFIA observations of [\ion{C}{2}]}

\shorttitle{Extraplanar [C II] with SOFIA}

\author[0000-0001-8362-4094]{William T. Reach}
\affil{Universities Space Research Association, MS 232-11, Moffett Field, CA 94035, USA}
\email{wreach@sofia.usra.edu}

\author{Dario Fadda}
\affil{Universities Space Research Association, MS 232-11, Moffett Field, CA 94035, USA}
\email{dfadda@sofia.usra.edu}

\author{Richard J. Rand}
\affil{Department of Physics and Astronomy, University of New Mexico, 800 Yale Blvd, NE, Albuquerque, NM 87131, USA }
\email{rjr@umn.edu}

\author{Gordon J. Stacey}
\affil{Astronomy Department, Cornell University, Ithaca, NY 14853}
\email{stacey@astro.cornell.edu}

\begin{abstract}

%Active star formation in the midplane of a galaxy, and collisions with other galaxies, can increase the thickness of disks. 
%It is thought that 
Bursts of localized star formation
 in galaxies can levitate material from the midplane. 
Spiral galaxies that are edge-on allow clear distinction of material that is levitated off the galaxies’ midplanes. 
We used SOFIA to
measure the vertical distribution of [\ion{C}{2}] 157.7 $\mu$m line emission for two nearby,
edge-on galaxies, NGC 891 and NGC 5907. We find that for the central region and actively-star-forming regions in the northern portion of NGC 891, and for NGC 5907, a thin ($0.3$ kpc) disk is
supplemented by a
thick disk with an exponential scale height of $\sim 2$ kpc.
The [\ion{C}{2}] is far more extended than mid-infrared emission
(0.1 kpc, tracing present-day massive star formation) but not as extended
as the \ion{H}{1} (100 kpc, tracing low-metallicity circum/inter-galactic matter).
The extraplanar [\ion{C}{2}] may arise in walls of chimneys
that connect the disk to the halo.

\end{abstract}

\section{Introduction}

The vertical distribution of material in disk galaxies is determined by a balance between gravity from the stellar disk and pressure (thermal, magnetic, relativistic particles) in the interstellar gas. In the Milky Way, cloud pressures and density of starlight can be measured locally, but without large-scale context due to our location within the disk. In particular, material far from the midplane of the galaxy is difficult to separate from local gas. 
Active star formation in the midplane of a galaxy, and collisions with other galaxies, can increase the
thickness of disks, {\bfc with stars (collisionless) and gas (collisional) responding
differently}. 
Bursts of localized star formation can form large-scale bubbles \citep{heiles90,breitschwerdt06} or drive material into the halo as `worms' \citep{heiles96}, whence it returns to
the midplane as a `galactic fountain' \citep{bregman80,norman89} or if sufficiently powerful escapes as a wind \citep{heckman90}.

\def\extra{
Galaxies are composed of dark matter, stars, and interstellar gas; in the Solar Neighborhood, these constituents comprise 42\%, 31\%, and 14\% of the mass surface density, respectively \citep{mckee15}. That the densities are in the same order of magnitude suggests that the constituents have significant mutual influence. 
A landmark survey of edge-on galaxies showed that the scale height of stars was independent of position along the major axes of galaxies, which would occur during galaxy evolution if there are continual random gravitational accelerations by spiral density waves and giant molecular clouds \citep{vanderkruit82}. 
The scale height of CO-traced molecular gas in the Milky Way was later observed to increase monotonically with galactocentric radius \citep{malhotra94,yim20}. 
The apparent discrepancy {\bfc between the radial dependence of scale heights}
may have been resolved once extinction was taken into account, and
the scale height stars, like CO, may actually increase with galactocentric distance
\citep{degrijs97}.
%, but these studies show one way interstellar scale heights relate to galaxy evolution. 
The existence of thick disks of stars remains an active topic \citep{comeron18}. 
}

Edge-on galaxies offer a prime opportunity to measure the vertical distribution of both stars and interstellar material.
For a galaxy that is exactly edge on, the angular distance from the ridge of peak brightness is readily interpreted as vertical displacement from the galactic midplane. 
There are several well-known, nearby, nearly edge-on galaxies, which exhibit distinct, dramatic dust lanes due to absorption of starlight by dust along a long path length through the disk. 
If a galaxy is inclined by at least 87$^\circ$ and is comparable in size to the Milky Way, then {\bfc the thin disk can be separated from potentially thicker distribution of 
`extraplanar' gas}. 
Fortunately, nature provides several such nearby, nearly-edge-on galaxies. 
Extraplanar gas extending to 5 kpc was detected in H$\alpha$ observations of the nearly edge-on galaxy NGC 891 \citep{rand97}, and even 13 kpc above NGC 5775 \citep{boettcher19,rand00}
requiring processes both to levitate the material and keep it ionized.
In another edge-on galaxy, NGC 5907, deep optical images revealed streams interpreted as the `ghost' of
dwarf galaxy collisions during formation \citep{martinez08}, although that galaxy 
shows no evidence for extraplanar diffuse ionized gas \citep{rand96}.

The scale height of the ISM in the Milky Way was measured from [\ion{C}{2}] observations with {\it Herschel}  \citet{langer14b}, with a FWHM of 172 pc, intermediate between molecular gas traced by CO (110 pc) and atomic gas traced by \ion{H}{1} (230 pc).  [\ion{C}{2}] emission can be produced by a wide range of interstellar gas, including ionized, atomic, and molecular. It is the most intense spectral line
from  many galaxies and dominates the cooling for  atomic and diffuse molecular gas. 
High-latitude clouds evidently contain a significant fraction of `dark gas,’ which is likely 
predominantly H$_2$ but without discernible emission from detectable tracers like CO.
The amount of such `dark gas’ was measured independently using gamma-rays as a total nucleon tracer (via the interaction between cosmic rays and interstellar protons \citep{grenier05} and far-infrared dust emission 
\citep{hrk88,reach15,leroy11}
as total column density tracers.
`Dark’ gas appears to be of order 40\% of the total molecular gas mas in the local Milky Way. That same proportion is unlikely to apply in all galaxies nor at all galactocentric radii nor distances from the midplane. 
Far-infrared observations of [\ion{C}{2}] 
can explore the presence of `dark’ gas in galaxy halos. 

Nearby galaxies provide the bridge between the Milky Way and distant galaxies, and nearby, edge-on galaxies provide the best access to the vertical distribution and the connection between galaxies' disks and halos.
For distant galaxies, [\ion{C}{2}] is one of the few spectral lines that can be used as a 
signpost for star formation, and a measure of local far-UV radiation field strength, hence star 
formation intensity  \citep{stacey91,pineda18}. 
In edge-on galaxies,
extraplanar, diffuse ionized gas, traced by H$\alpha$,  is more prominent with higher SFR surface density \citep[e.g.][]{rand96,rossadettmar03}, as is extraplanar dust absorption \citep{howk99}. 
The H$\alpha$ plumes extended from the midplane are possible chimneys where material that has been heated and accelerated
by active star formation regions vents into the halo.
For the best-studied edge-on galaxy, NGC 891,
the \ion{H}{1} distribution is very extended, with a thin disk of exponential scale height of 0.2 kpc and an extended distribution with approximate scale height of 2.5 kpc \citep{oosterloo07}. 
In a more face-on galaxy, NGC 6946, the \ion{H}{1} distribution reveals
in-falling gas that may lead to holes in midplane \citep{boosma08}, so the extended extraplanar gas can not only be the up-welling exhaust of chimneys from massive stars but also the returning gas. 
Extraplanar \ion{H}{1} was found to have a global infall rate of $\sim 25$ \kms, which could be the
incoming part of galactic fountain \citep{heald11,marasco19}.
A thick disk with scale height 1.4 kpc was detected in far-infrared dust observations of NGC 891 \citep{bocchio16}.
Another component of galaxies, relativistic particles,
has an extended scale height, measured at 1.3 kpc for the disk of NGC 891 \citep{changes19}.
In [\ion{C}{2}], there was a hint of an extended disk already in the NGC 891 {\it Herschel} observations with a measured scale height of 300 pc for the inner 3 kpc radial distance that was observed
\citep{hughes15}. 
%The trends for CO are not well known.

In this paper we address the [\ion{C}{2}] distribution,  tracing the disk-halo connection, for two prominent, nearby, edge-on galaxies NGC 891 and NGC 5907, to measure the vertical distribution of a tracer of higher-density material than is traced by H$\alpha$, \ion{H}{1}, or cosmic
rays. 

\section{Observations\label{sec:observations}}

The observations were part of SOFIA Cycle 6 observing program 06\_0010.
The Field Imaging Far-Infrared Line Spectrometer (FIFI-LS; \citet{fischer18}) was used to map the vertical extents of NGC 891 and NGC 5907 in the [\ion{C}{2}] 157.741 $\mu$m (rest frame) line. 
The spectral resolving power of FIFI-LS in the [\ion{C}{2}] line is 1200, corresponding to velocity resolution 250 km~s$^{-1}$.
Because FIFI-LS is a dual-channel instrument, we simultaneously observed the [\ion{O}{3}]
88.356 $\mu$m line; the signal-to-noise was not sufficient in that line, so those data will not be discussed in this paper.
Table~\ref{obstab} shows the SOFIA observation conditions, which were conducted using 5.8 hr of flight time.

\def\tnm{\tablenotemark}
\begin{deluxetable*}{ccccccccc}
%\tabletypesize{\scriptsize} 
%\tablewidth{0pt}
%\tablenum{text}
\tablecolumns{6}
\tablecaption{SOFIA Observing Log\label{obstab}} 
\tablehead{
\colhead{Target} & \colhead{Date} & \colhead{Flight} & \colhead{Altitude} & \colhead{Elevation} & \colhead{Exposure} \\
\colhead{} & \colhead{} & \colhead{} & \colhead{(feet)} & \colhead{($^\circ$)} & \colhead{(min)} \\
}
\startdata
NGC 891-South & 2019 Feb 28 & 549 & 38,500 & 47 & 34\\
%NGC 4244  & 2019 Mar 01 & 550 & RTB    & & 0\\
NGC 5907  & 2019 Mar 02 & 551 & 42,800 & 45 & 32\\
NGC 891-North & 2019 Nov 05 & 635 & 44,380 & 49 & 17\\
NGC 891-North & 2019 Nov 08 & 638 & 42,280 & 53 & 22\\
NGC 891-North & 2019 Nov 13 & 639 & 45,890 & 52 & 16
\enddata
%\tablenotetext{a}{Exposure time per FIFI-LS field ($1'\times 1'$).}
\end{deluxetable*}

\begin{figure*}
\includegraphics[height=4.5in]{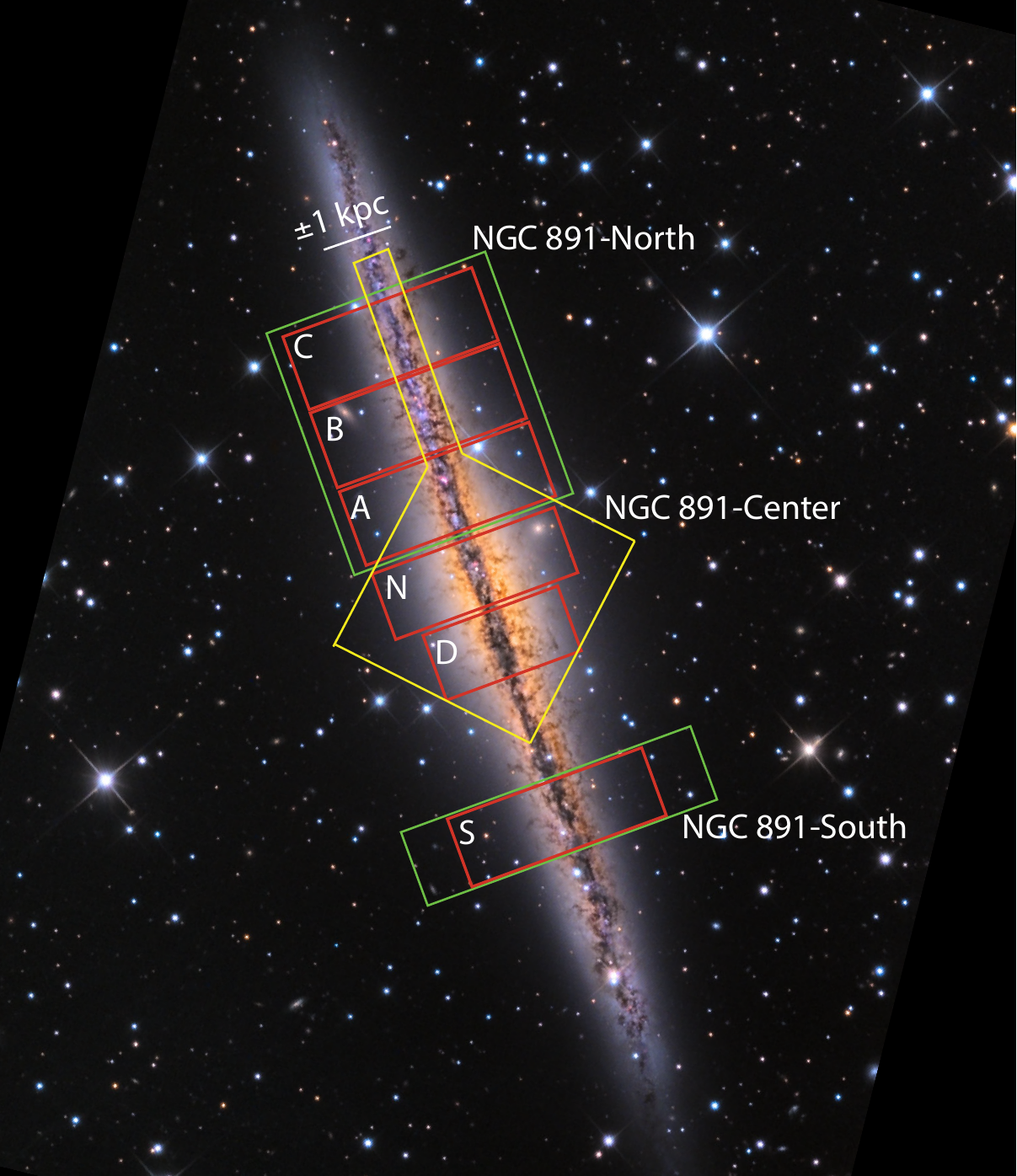}
\includegraphics[height=4.5in]{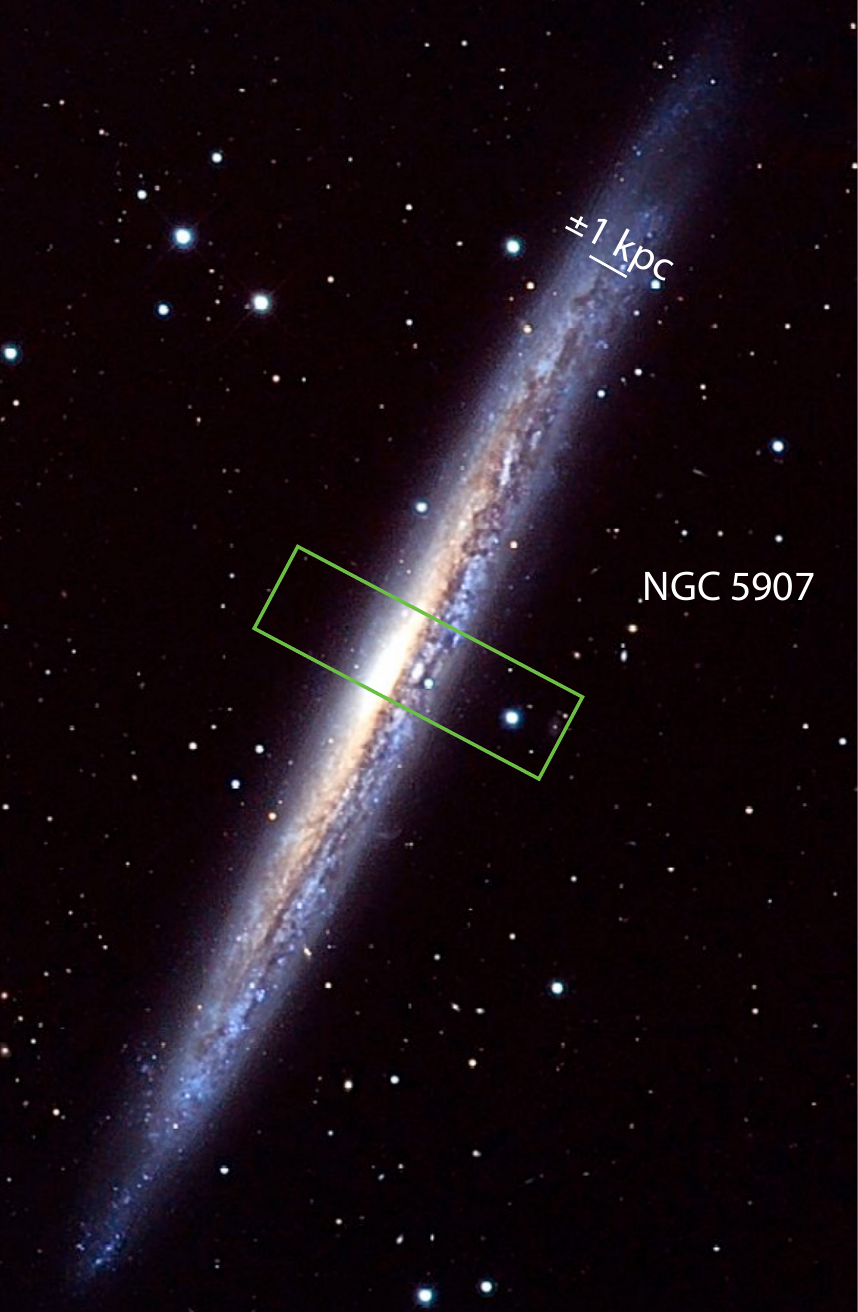}
    \caption{Overlays showing the locations of the fields in which [\ion{C}{2}] was observed with SOFIA for NGC~891 (left) and NGC~5907 (right) on visible-light images {\bfc (A. Block, Mt. Lemmon SkyCenter Schulman telescope, wikimedia commons).}
    The green boxes show the on-source coverage with SOFIA, and the yellow polygon outlines the
    region observed by {\it Herschel}. Labeled red rectangles show the strips from which 
    vertical profiles were extracted in Fig.~\ref{fig:prof}.
    Scale bars indicating $\pm$1 kpc ($52''$ for NGC 891 and $27''$ for NGC 5907) are shown for reference.
    \label{fig:overlay}}
\end{figure*}

Figure~\ref{fig:overlay} shows the regions that were imaged: two fields, NGC 891-North and NGC 891-South, 
cover a range of longitudes along the midplane and 3 kpc vertical extend from the midplane of that galaxy, and one field 
covers the central longitudes of NGC~5907. 
The observations were performed with the  secondary mirror chopping between the galaxy and a reference field, which alternated between fields on either side of the galaxy $300''$ from the midplane. The instantaneous field of view was $60"\times 60"$ with  $5\times 5$ spaxels of $12"\times 12"$ each.
The total field we covered comprise a set of dithered raster pointings that span a large portion of the galaxy while filling in the gaps between spaxels, 
which allows for drizzling onto a fine, well-sampled grid during data reduction.
The data were processed into data cubes (RA, Dec, Velocity) by the SOFIA science center including drizzling onto a fine grids of $5"$ sampling.
At the distances of NGC 891 and NGC 5907, the  pixel sizes are
190 pc and 370 pc, respectively.

To improve the separation between the thin, bright emission from the midplane and the 
extended, faint emission of the extraplanar gas, we reprocessed the data using a narrower
spatial kernel. Whereas the standard data processing uses a spatial window of $2\times$ the
beam and a wavelength kernel of $1.2\times$ the
spectral resolution.
We reprocessed using a spatial window of $1\times$ the
beam and a wavelength kernel of $0.5\times$ the
spectral resolution. This processing more closely matches that of {\it Herschel}/PACS, aiding
inter-comparison.

Archival, fully reduced, [\ion{C}{2}] line integral images of NGC 891, part of the  {\it Herschel} 
Very Nearby Galaxies Survey \citep{hughes15}, were analyzed to generate a vertical profile 
for the inner galaxy at 1--2 kpc galactocentric radii.
Inspecting Fig. 5 from \citet{hughes15}, there was already a hint of a change in the slope of surface brightness versus altitude, indicating an extraplanar gas component distinct from gas in the disk of NGC 891. 
The {\it Herschel} observations were unchopped and had to be corrected for transients \citep{fadda16}.
Comparing the {\it Herschel} and SOFIA observations in detail shows a potentially steeper
decline of [\ion{C}{2}] in the {\it Herschel} data from 0--1 kpc. Only a small part of this difference is the slightly larger SOFIA beam.

To verify the calibration and quality of the SOFIA/FIFI-LS data, we extracted the spectrum of a portion of the NGC 891 disk (the southernmost portion of the disk in the NGC 891-North field) that was observed with {\it Herschel}/PACS. We used the same spatial aperture and spectral baseline to extract the spectrum.
The in-aperture flux measured with FIFI-LS was $(2.78\pm 0.09) \times 10^{-15}$ W~m$^{-2}$,
while that measured with PACS was $(2.76\pm 0.01)\times 10^{-15}$ W~m$^{-2}$, in excellent agreement. 
To check the 
effect of using the FIFI-LS spectral baseline, we used a wider baseline in the PACS spectrum to
find a flux 4\% higher. A $\sim 4$\% uncertainty in flux due to bandpass differences between FIFI-LS and PACS have negligible effect on the analysis in this paper.

\section{Results}

\subsection{NGC 891}

\begin{figure*}
    \centering
%    \plottwo{FIFI-N-AB-spec.png}{PACS-N-spec.png}
    \plotone{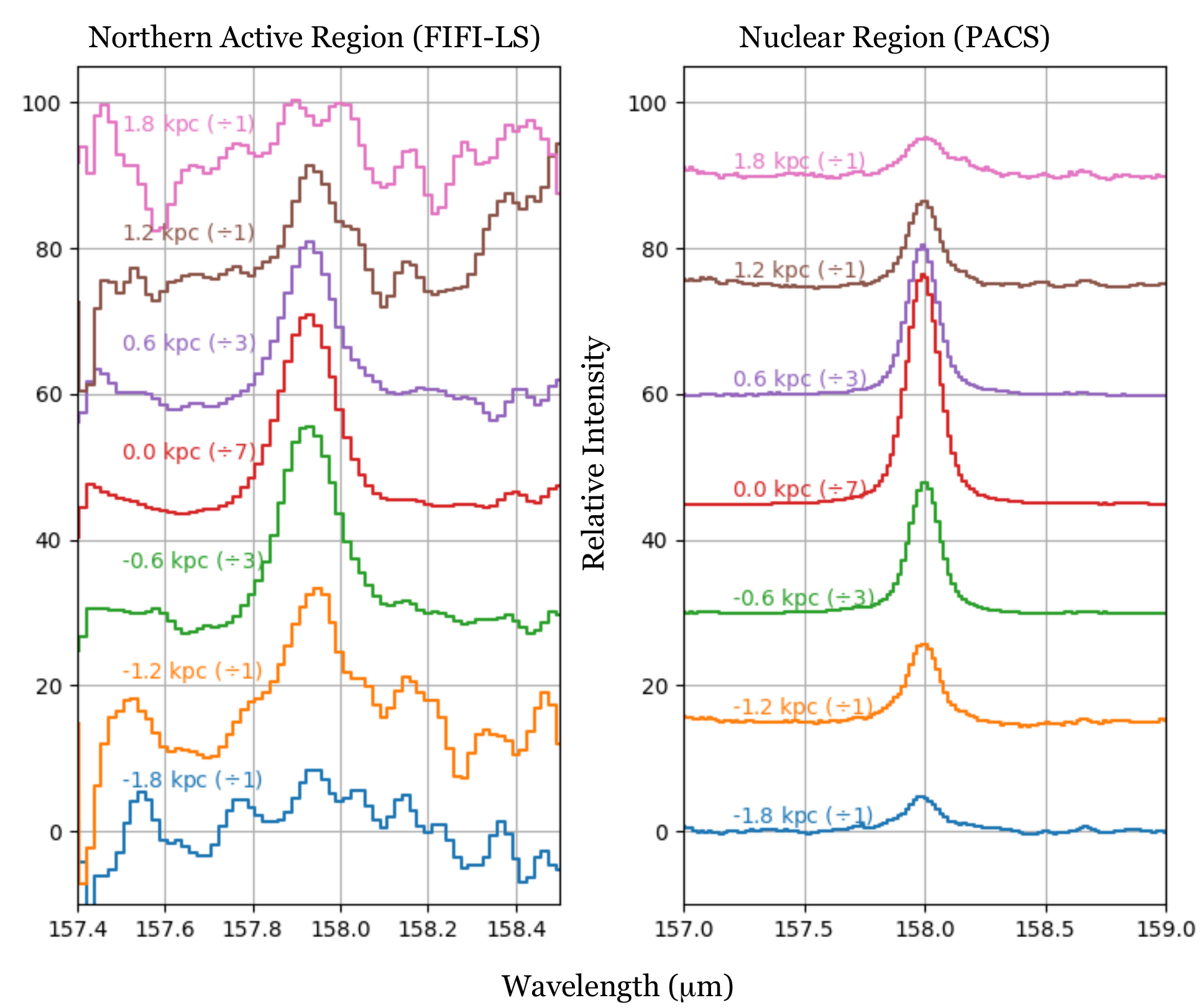}
    \caption{Spectra of NGC 891 observed by SOFIA in the northern disk (left panel; A \& B in Fig.~\ref{fig:overlay}) and by {\it Herschel} (right panel; N in Fig.~\ref{fig:overlay} through the nucleus) in the [\ion{C}{2}] 157.741 $\mu$m line. The spectra are averaged over  longitudes in the NGC891-North map, at each of 7 labeled distances (in kpc) from the midplane. 
    The spectra divided by scale factors (labeled) to better fit onto the diagram.
    \label{fig:spec}}
\end{figure*}

Figure~\ref{fig:spec} shows the averaged [\ion{C}{2}] spectra at 5 different altitudes, $z$, above the midplane, where the $+z$ direction is toward celestial northwest.
Each  spectrum was averaged over longitudes covering the bright midplane \ion{H}{2} regions in the Northern NGC 891-North field (see Fig.~\ref{fig:overlay}), and each spectrum includes  
a 1-beam (0.6 kpc) strip of altitudes centered on the labeled altitude. The midplane spectra are detected with a signal-to-noise of approximately 40 per wavelength resolution element. 
The spectra at $\pm1.2$ kpc clearly detect the [\ion{C}{2}] line.

\begin{figure}
    \centering
    \includegraphics[width=4in]{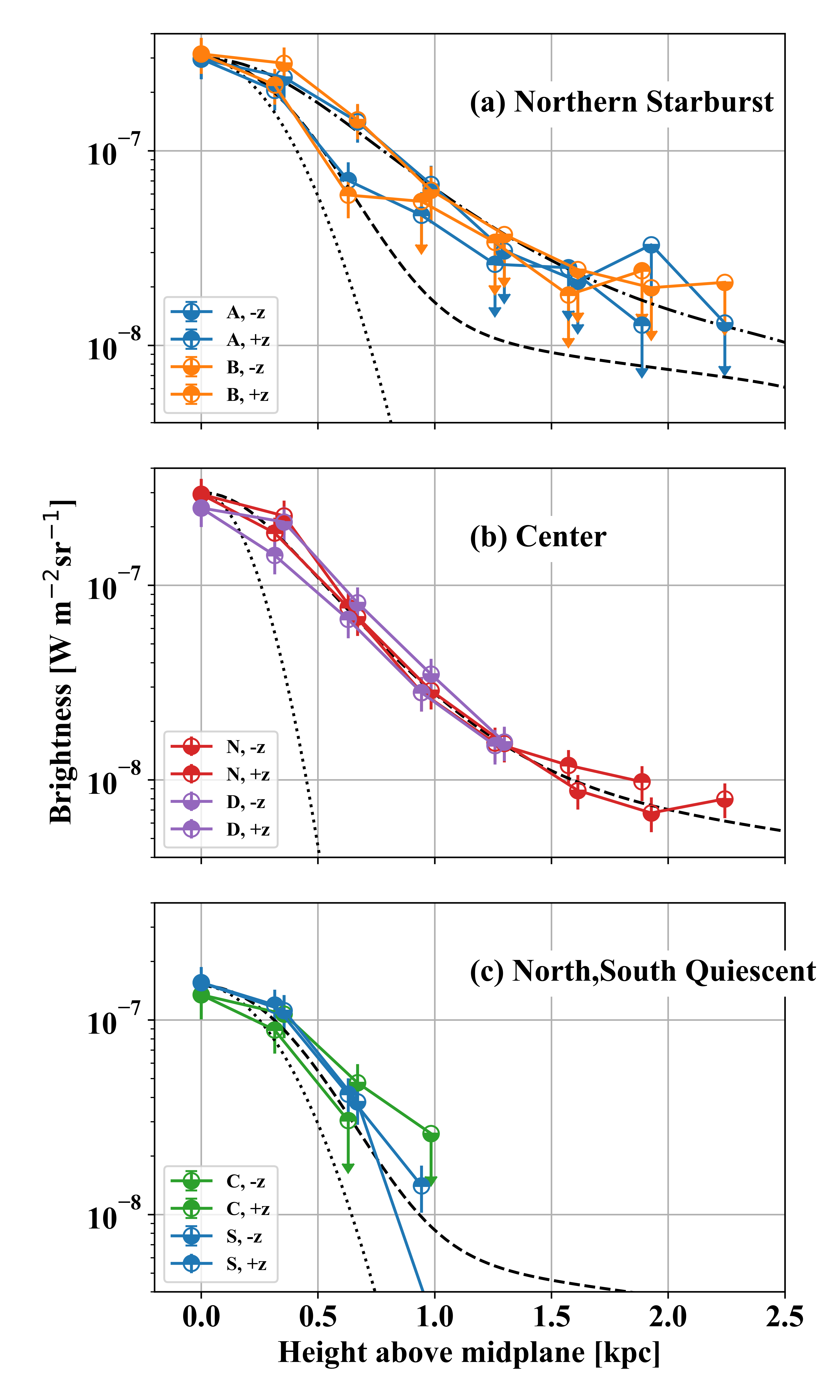}
    \caption{Vertical profiles of [\ion{C}{2}] in NGC 891, showing the line brightness versus height above the midplane for slices through the galaxy at 7 different longitudes. The location of each slice
    through the galaxy is shown in Fig.\ref{fig:overlay}. The top panel (a) shows the vertical profile at two locations in the Northern field observed by SOFIA/FIFI-LS. The profiles for the +$z$ and -$z$ directions are shown separately, with symbols with the upper and lower halves, respectively, filled. The two longitudes in this panel cross through bright \ion{H}{2} regions in the midplane. The center panel (b) shows profiles at through locations closer to the center of the galaxy observed by Herschel/PACS.
    The bottom panel (c) shows profiles for the southern field and a northern field observed by SOFIA/FIFI-LS where the star formation activity in the midplane is lower than in panel (a).
    In each panel, the black dotted line shows the observatory point spread function (PSF), normalized to the midplane brightness.
    Also in each panel, dashed black lines show exponential scale-height models (convolved with the PSF) described in the text and listed in Table~\ref{fittab}. In panel (a), two such fits are shown, for the -$z$ (dashed-dot) and minimum fit to +$z$ (dashed) directions.
    \label{fig:prof}}
\end{figure}

Figure~\ref{fig:prof} shows the derived vertical distribution of [\ion{C}{2}] for NGC 891 at 7 different longitudes, at altitudes
separated by approximately 1 beam. 
For NGC 891-N, we estimated the signal-to-noise at each $z$ from the fluctuations in the longitude-averaged spectra 
within the spectral baseline windows. 
The midplane [\ion{C}{2}] flux is detected with signal-to-noise greater than 100, and the flux at $\pm1$ kpc is detected with signal-to-noise greater than 10. The flux at $\pm2$ kpc remains above
the detection limit with signal-to-noise approximately 4, 
but the signal is not detected beyond 2.4 kpc from the midplane.

To quantify the vertical extent of the [\ion{C}{2}] emission, we compare the observed vertical profiles to
those expected for exponential disks,
$I\propto \exp{(-|z|/H})$,
convolved with the angular resolution of SOFIA/FIFI-LS ($16''$) or Herschel/PACS ($10''$).
{\bfc We first fit a single-scale-height model, which was adequate for the NGC 891-S profile. Then we fit a two-scale-height model for the NGC 891-N and nuclear profiles. 
The two-scale-height model includes coupled (partially degenerate) parameters and is not
a unique solution; in particular, 
the second scale
height is not tightly constrained due to lack of signal-to-noise at the highest $z$.}
The convolved models are show in Fig.~\ref{fig:prof} as dashed lines and reasonably well describe
the overall trend. The model scale heights are summarized in Table~\ref{fittab}. 
The discrepancies {\bfc between model and observed profiles} are partly due to real structure in longitude (discussed below). 
{\it bfc orientationFurthermore, the northern and southern profiles are different, as can be 
seen in Fig.~\ref{fig:prof}, indicating asymmetry in the origin of the 
extraplanar material.}

\begin{deluxetable*}{lcccc}
\tablecolumns{5}
\tablecaption{Scale Height Fits\label{fittab}} 
\tablehead{
\colhead{} & \colhead{NGC 891-N, starburst, -$z$} & \colhead{NGC 891-center} & \colhead{NGC 891-S} & \colhead{NGC 5907}    }
\startdata
resolution\tablenotemark{a} (kpc)           & 0.32  &  0.20  & 0.32  & 0.63  \\
scale height (kpc): thinner & $0.40\pm 0.1$  & $0.5\pm 0.1$ & $0.3\pm 0.08$   & $0.37\pm 0.05$ \\
scale height (kpc): thick   &  $2.8\pm 0.7$  & $2.8\pm 0.7$  & ...             & $\sim 3$ \\
\enddata
\tablenotetext{a}{Half-width at half-maximum. Observed with SOFIA except NGC891-center, with {\it Herschel}}
\end{deluxetable*}

At low altitudes from the NGC 891 midplane, the brightness is primarily from a thin disk with scale height in the range 
0.25--0.4 kpc.
It is thickest on the northern field, where there are giant \ion{H}{2} regions in the midplane,
and thinner in the more-quiescent southern field.
For the active starburst portions of the northern field (in the -$z$ direction) and the central part of the galaxy (both directions), 
a second component with scale height around 2.8 kpc is required.
This wider component contributes only 3\% of the [\ion{C}{2}] model brightness in the midplane (before convolution with
the PSF), but with its larger scale height it contributes 22\% of the total brightness integrated over $z$ and most
of the brightness for lines of sight above 1.2 kpc from the midplane.

\subsection{NGC 5907}

\begin{figure*}
    \centering
    \plotone{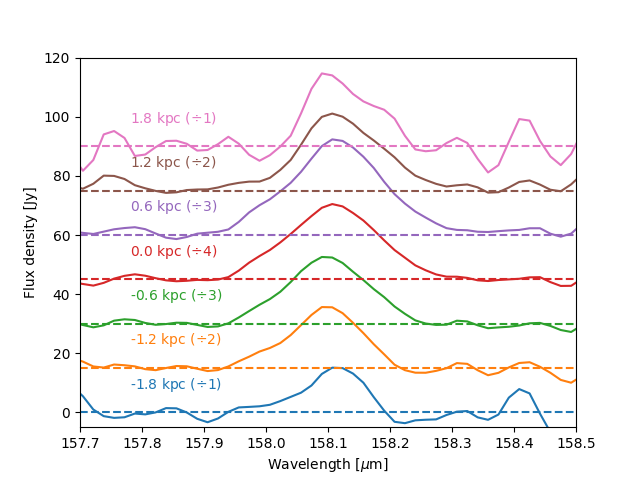}
    \caption{Spectra of NGC 5907 observed by SOFIA  in the [\ion{C}{2}] 157.741 $\mu$m line averaged over all longitudes 
    observed, at each of 7 labeled distances (in kpc) from the midplane. 
    The spectra divided by scale factors (labeled) to better fit onto the diagram.
    \label{fig:spec5907}}
\end{figure*}

\begin{figure}
    \centering
    \includegraphics[width=4in]{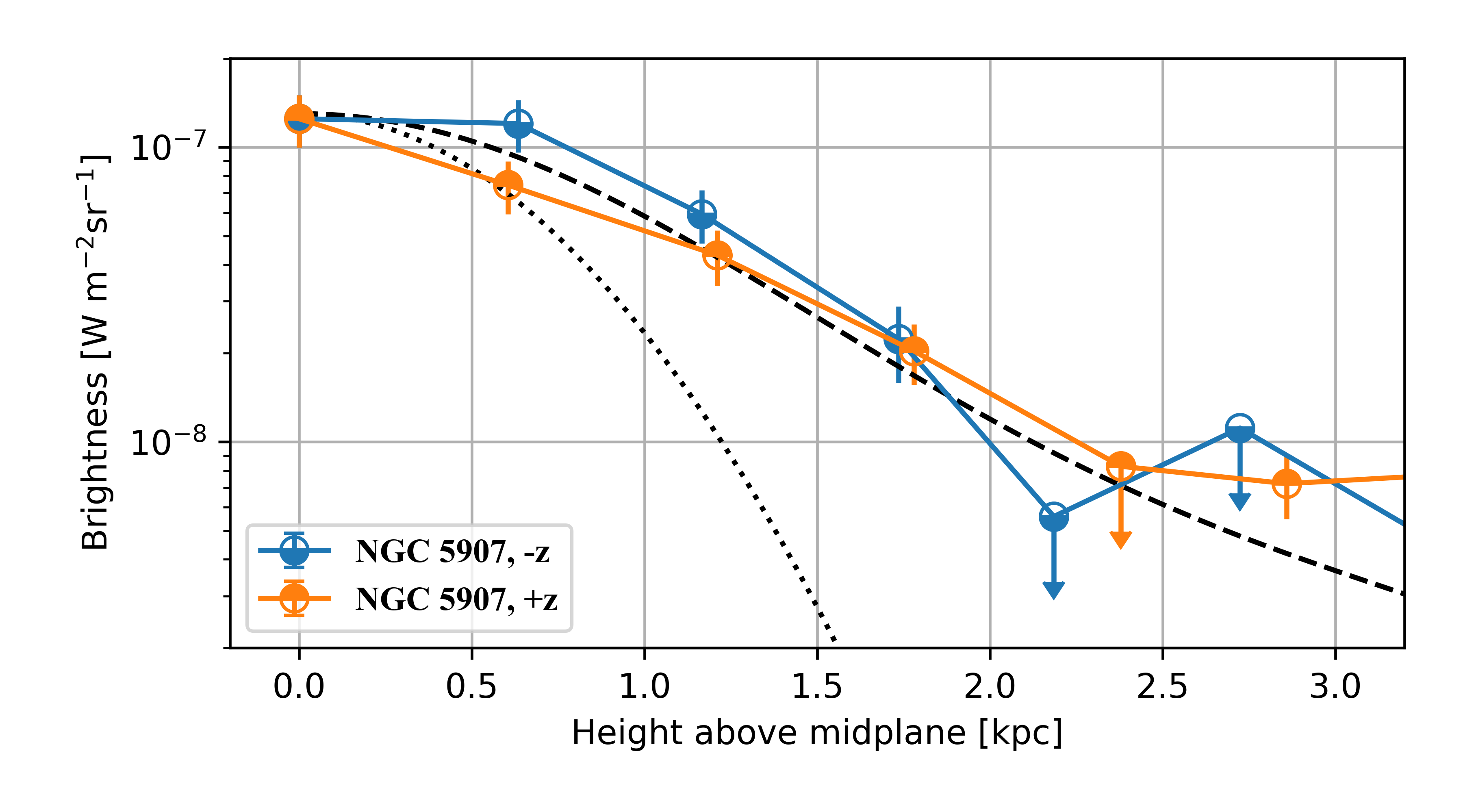}
    \caption{Vertical profiles of [\ion{C}{2}] in NGC 5907, showing the line brightness versus height above the midplane for slices through the center of the galaxy. The profiles for the +$z$ and -$z$ directions are shown separately, with symbols with the upper and lower halves, respectively, filled. The black dotted line shows the observatory point spread function (PSF), normalized to the midplane brightness.
    The dashed black lines show  an exponential scale-height models (convolved with the PSF) described in the text and listed in Table~\ref{fittab}. 
    \label{prof5907}}
\end{figure}

{\bfc Figure~\ref{fig:spec5907} shows the averaged [\ion{C}{2}] spectra at 7 different altitudes for NGC 5907, with the $+z$ direction is toward celestial northeast.}
In NGC 5907, the {\bfc width} of the 
[\ion{C}{2}] is  much greater than an unresolved disk, despite the lower physical resolution of the observations {\bfc compared to NGC 891}
(due to the greater distance of {\bfc NGC 5907}). 
Figure~\ref{fig:spec5907} shows clear
detections of [\ion{C}{2}] out to 1.8 kpc from the midplane.
Figure~\ref{prof5907} shows the vertical profile compared to the convolved exponential
disk model as used in the previous subsection. The illustrated model does not match the data particularly well. It does, however, show that the vertical distributions
on the two sides (+$z$ and -$z$) of the midplane are different, with the galaxy somewhat thicker in the -$z$ direction near the midplane, but having a longer `tail' to higher $z$ in the +$z$ direction.
The SOFIA observations detect the extended extraplanar gas up to 1.8 kpc from the midplane in both directions.

The measured scale heights of the galaxies are somewhat larger than the true scale heights,
because the galaxies are not seen completely edge-on.
To estimate this effect for NGC~5907, we modeled a tilted disk that is exponential in galactocentric radius and absolute vertical extent, then compared the profiles at a range of inclinations.
For $i=90^\circ$, perfectly edge-on, the observed vertical profile is the same as the projected. As $i$ decreases, the
predicted profile for the double exponential disk model is closely matched by a smoothed version of the true vertical profile. 
If the radial exponential scale length is $H$, then 
to a reasonable approximation, the observed vertical tilted profile is the true vertical profile smoothed by 16 $H \cos i$.
For NGC~5907, at $i=86.5^\circ$ \citep{garcia97}, the observed profile is the true one smoothed by 0.98 $H$, so the true scale
height is approximately $0.7 / \sqrt{1+0.98^2}=0.5$ kpc. Note that this approximation only applies
for smooth, exponential disk; some of the apparent vertical variation may be due to arms or 
far-outer-galaxy \ion{H}{2} region 
(though this should make the profile very asymmetric which is not observed).
{\bfc (In comparison, for NGC~891, at $i>89^\circ$ \citep{oosterloo07}, the true profile only needs to be smoothed by $0.28 H$ or less,
so the observed scale height is less than 4\% larger than the true one.)}

\section{Discussion}

\subsection{Relationship between extraplanar [C II] and Chimneys}

The presence of metal-bearing material far from the midplane of the galaxies
suggests a cycling of material such as in a galactic fountain, where massive
star formation expels material from the midplane
\citep[][]{shapirofield76,bregman80,norman89}.
The extraplanar gas must be either maintained at its altitude by a significant pressure of underlying gas, or 
 displaced from the midplane 
by transient events after which it returns to the midplane.
Vertical extensions in edge-on galaxies have been interpreted as 
`chimneys' through which massive star-forming regions vent material into the halo \citep{rand90}; such structures in the Milky Way  have been 
labeled  `worms' \citep{heiles96}.

\begin{figure}
    \centering
    \includegraphics[scale=.5]{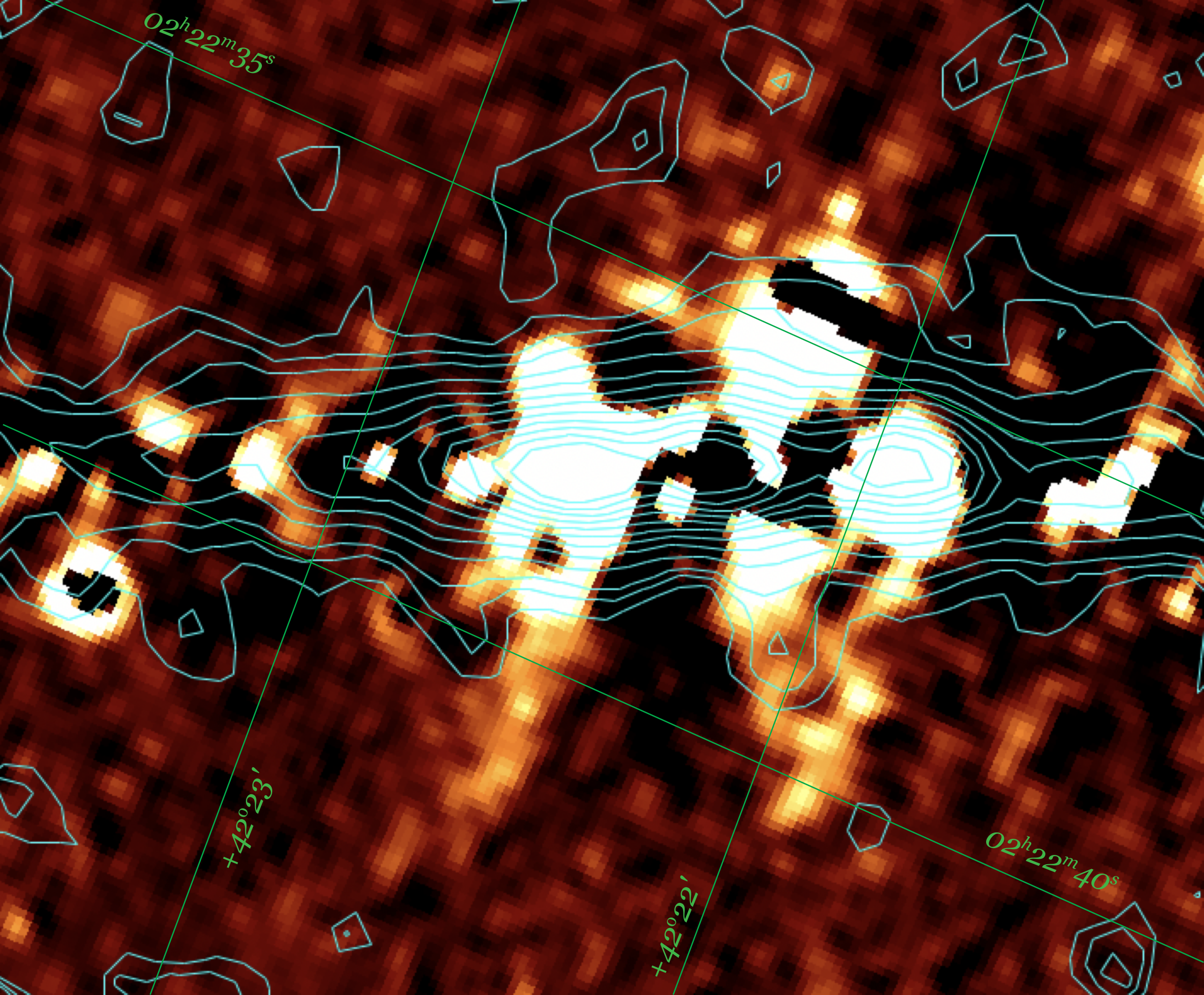}
    \caption{Overlay of SOFIA [\ion{C}{2}] emission contours on the spatial-filtered H$\alpha$ image \citep[from][]{rand90}. The region covered is shown in 
    Fig.~\ref{fig:overlay} as the northern box; it has been rotated counterclockwise so that the midplane is horizontal. The [\ion{C}{2}] brightness is from a single 
    channel at -200 \kms\ relative to the central velocity of the entire galaxy. 
    The H$\alpha$ image shows the giant \ion{H}{2} regions of the midplane (partially interrupted by extinction lanes), with prominent vertical extensions that
    correspond to [\ion{C}{2}] features discussed in the text. {\bfc The coordinate grid (J2000) is overlaid.}
    \label{ciiha}}
\end{figure}

The new [\ion{C}{2}] images show that ionized carbon also extends to the vertical heights of the chimneys at 2 kpc. Our view of the 2-dimensional distribution of the 
[\ion{C}{2}] is limited by signal to noise, but for NGC 891 some reasonably 
well-defined structure is present. Figure~\ref{ciiha} shows a direct overlay
of the H$\alpha$ on the [\ion{C}{2}] image. The H$\alpha$ image has already been spatially filtered to emphasize structure. Six giant \ion{H}{2} regions in the midplane are evident, with the primary ones near the center of the imaged
region. The region containing the \ion{H}{2} regions is also prominent in
[\ion{C}{2}]. It is likely a combination of molecular clouds, massive stars and the interfaces between that are photodissociation regions, which are likely the source of most of the [\ion{C}{2}] in NGC~891 \citep{stacey10}. 

Extending vertically downward from the brightest \ion{H}{2} regions {\bfc in Figure~\ref{ciiha}}
are two
roughly vertical H$\alpha$ features. We will call them `chimneys', 
for descriptive purposes. 
{\bfc There are two prominent H$\alpha$ features, and both have associated [\ion{C}{2}].
These features could conceivably be part of the same structure (i.e. limb-brightened walls of one chimney); however, that structure would be 2 kpc across and would likely be 
highly distorted by galactic shear in 60 Myr (if the rotational shear of NGC 891 is similar
to that of the Milky Way). 
We will consider these structures to be two separate features, most likely
related to the \ion{H}{2} regions from which they appear to extend.

The H$\alpha$ plume extended downward from the midplane, just left of center
in Fig.~\ref{ciiha}, has a somewhat diagonal [\ion{C}{2}] feature that is centered
in the base of the H$\alpha$, filling in a location where the H$\alpha$ appears to
have a void. This configuration could be caused by denser material near the midplane that 
remains relatively neutral (most of the H in atomic or molecular form) and also by
extinction blocking the H$\alpha$.

For the other prominent downward chimney, just right of center in Fig.~\ref{ciiha},
there is again a corresponding [\ion{C}{2}]
feature in the `base'. The base of this chimney is also evident in a 
{\it Hubble} image as a filament extending about 690 pc from the midplane \citep[Fig. 8 of][]{rossa04}.

On the upward side of the midplane, there are no prominent H$\alpha$ chimneys in this part of the galaxy. The H$\alpha$ image shows one prominent vertical feature that lies just to the right of Fig.~\ref{ciiha} (outside the area well-covered  by SOFIA) and near the eastern edge of the PACS image, making it difficult to assfess the correspondence.}

With the present data, it cannot be unequivocally determined whether the [\ion{C}{2}] and
H$\alpha$ structures are physically related. However, the correspondence
is strong enough that we explore the possible implications. 
The [\ion{C}{2}] and H$\alpha$ lines can trace regions of different 
physical conditions and, most importantly, of different illumination
by hot, ionizing stars in the midplane. The massive star forming clusters 
that may have ejected the material into the halo (or a previous generation
in the same spiral arm) send ionizing photons on straight sightlines wherever
they avoid intercepting optically thick clouds. Regions that are behind
an H column density greater than $10^{21}$ cm$^{-2}$ have enough intervening
dust to be shielded from far-ultraviolet photons so the H is neutral. 
The chimneys will therefore have `walls' that will not necessarily emit 
in H$\alpha$. The [\ion{C}{2}], then, traces the colder gas in the walls
of the chimneys.

Assuming the [\ion{C}{2}] line emission arises from neutral gas, we use the observed surface brightness of $1.9\times 10^{-4}$ erg~cm$^2$~s$^{-1}$~sr$^{-1}$ to derive the neutral gas column density.  
Using line cooling rate from \citet{goldsmith12} and carbon abundance from \citet{cardelli96}, the observed [\ion{C}{2}] 
requires a significant column density, of order 
\begin{equation}
N_{\rm H} \simeq 4\times 10^{22} f(T)^{-\frac{1}{2}} \left(\frac{L}{100\,{\rm pc}}\right)^\frac{1}{2} {\rm cm}^{-2} 
\end{equation}
where the emitting region is has  a path length of 100 pc.
Equivalently, this column density estimate applies if the emitting region has a 
volume density of 65 cm$^{-3}$, and it scales inversely with volume density. 
The temperature dependence is weak (unless the gas is very cold):
\begin{equation}
f(T)\equiv \exp^{-0.91\left(\frac{100}{T}-1\right)} \left(\frac{T}{100}\right)^{0.14}
\end{equation}
which has values $f(T)$= 0.1, 0.4, 1, 1.9, and 2.6 at $T$=30, 50, 100, and 500 K, respectively.
The high column density means that the [\ion{C}{2}] emitting region is optically thick to 
UV and visible radiation. Ionization of C must be maintained by optically
thin path lengths interior to the chimney walls, which then form a photodissociation region.
Dust mixed with
the ionized gas in the H$\alpha$ emitting interior of the chimney may provide
some opacity, 
but the dust/gas ratio could be low due to
dust destruction during the violent ejection of material from the midplane.

{\bfc
If, on the other had, the [\ion{C}{2}] arises from ionized gas, and the emitting region  has path length $L$, then the  emission measure of ionized gas is
\begin{equation}
{\rm EM} \equiv \int n_e^2 dL = 1.5\times 10^3 \left(\frac{T}{10^4}\right)^{0.35} \,\, {\rm cm}^{-6}{\rm pc}
\end{equation}
for the  observed [\ion{C}{2}] surface brightness and Milky Way metalicity \citep{reynolds92}.
The implied column density is significantly reduced (by a factor of $\sim 40$) if the gas is ionized, 
because excitation by electrons, much more abundant in fully ionized gas than in a PDR,
is more efficient. 
Keeping the gas ionized requires a significant production of photons sufficiently energetic
to ionize H. To balance the total recombination rate of the region, the ionizing photon
production rate must be:
\begin{equation}
R_{\rm ion} \sim 5\times 10^{49} \left(\frac{L}{100\,{\rm pc}}\right)^\frac{1}{2} \,\, {\rm s}^{-1}.
\end{equation}
An ionization rate this high can be maintained if 2\% of the ionizing photons from an open cluster of mass $10^5 M_\odot$
with largest star of 50 $M_\odot$ escapes \citep{sternberg03}.
The giant \ion{H}{2} regions in the midplane underneath the H$\alpha$ chimneys may well be able to supply the needed ionizing photons.
Observations of  many giant \ion{H}{2} regions in other galaxies have sufficient photon production even with 1\% escape rates, as does W49 and NGC 3603 in the Milky Way, and 30 Dor in the LMC and  N66 in the SMC \citep{kennicut84}. It thus appears plausible that the [\ion{C}{2}] could arise from a large amount of
neutral gas in PDRs in chimney walls, or from a smaller amount of gas kept fully ionized by ultraviolet
photons leaking from the giant \ion{H}{2} regions in the midplane.
}

The new observations presented here show some evidence for structure of the extraplanar gas, but with improved sensitivity the issue could be resolved and address the structure of the disk-halo interface in more detail.
The {\it Herschel} observations of NGC 891 included a square map of the central portion of the galaxy and a strip along the midplane. The strip does not extend sufficient{\bfc ly}  to trace the extended gas. The 
square map of the central portion of NGC 891 extends to partially overlap about 15\% of our NGC 891-North SOFIA field  our observation, but due to its orientation, the overlap is primarily near the midplane, which does not allow for meaningful comparison or search for [\ion{C}{2}] associated with chimneys apart from the inner portion
of the galaxy. We see no clear evidence for [\ion{C}{2}] structures 
associated with H$\alpha$ plumes that extend above and below
the galaxy's nucleus and above a giant \ion{H}{2} region in the midplane north of the nucleus. Figure~\ref{fig:ngc891pah} shows the vertical profile for the central portion of NGC~891 (measured with {\it Herschel}
together with the vertical profile for the northern portion of the disk (measured with SOFIA).
The central portion of the galaxy images shows a steady
decrease in [\ion{C}{2}] with distance from the midplane indicating the 
cold, extraplanar gas above the inner portion of the galaxy is dominated
by the thin disk, with a lower contribution from the thick disk.

\subsection{Comparison of scale height with other tracers}

\begin{figure*}
    \centering
    \includegraphics[width=4in]{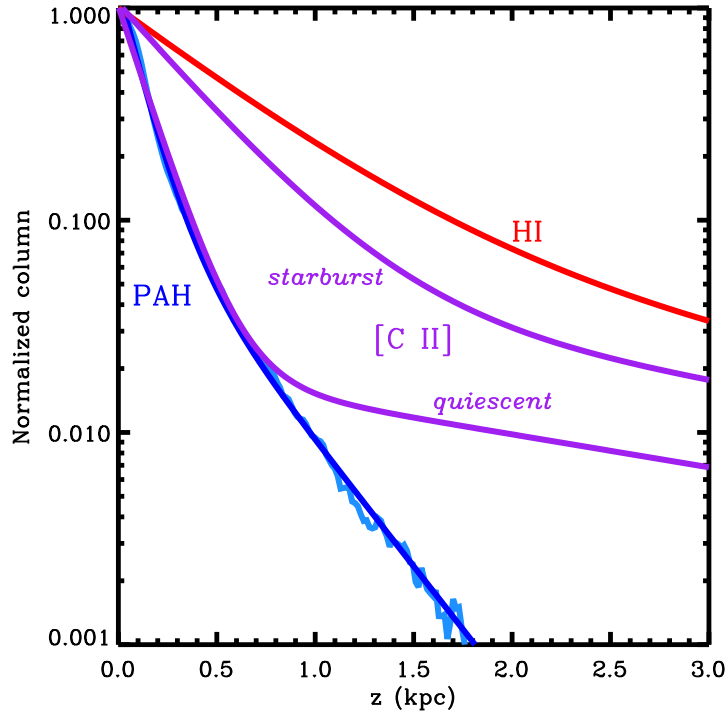}
    \caption{Vertical profiles of NGC 891. The \ion{H}{1} (red; {\bfc resolution 1.3 kpc}) extends well beyond the edge of the figure and includes extensive circumgalactic gas likely due to interactions with nearby galaxies. The PAH (blue; {\bfc resolution 0.2 kpc}) emission arises closer to the midplane and traces the present location of massive star formation. 
    The [\ion{C}{2}] distribution (purple; {\bfc resolution 0.7 kpc}) is shown for the starburst and quiescent locations from
    Fig.~\ref{fig:prof}. The quiescent profile is primarily consistent with the profile of present star formation, traced by PAH, while the more extended profile for the starburst (regions A+B  labeled 
    on the image in Fig.~\ref{fig:overlay}) has a more significant extended component.
    \label{fig:ngc891pah}}
\end{figure*}

Figure~\ref{fig:ngc891pah} compares the vertical profiles of various tracers of the ISM in NGC 891.
The \ion{H}{1} profile is from radio observations of the 21-cm line with the Westerbork Radio Synthesis Telescope \citep{oosterloo07}.
The polycyclic aromatic hydrocarbon (PAH) profile was measured from archival mid-infrared observations 
with {\it Spitzer} in a broad waveband centered at 8 $\mu$m dominated by PAH emission. The same galactocentric radius range (3--7 kpc) as the SOFIA observation was extracted.
The  differences between the profiles show that they trace 
different phenomena.
The widely distributed \ion{H}{1} extends out to 100 kpc,
with the inner portion potentially originating from a galactic fountain while the
most distant gas is likely  either primordial or stripped by gravitational
interactions with other galaxies \citep{das20}. 
The thin disk in  PAH emission that dominates the luminosity of the galaxy is due to present-day star formation {\bfc \citep{smith07pah}. }
{\bfc  The molecular gas scale height traced by CO is 240 pc \citep{scoville93,yim11}, similar to the PAH, representing the reservoir of material forming stars.}

The [\ion{C}{2}] and H$\alpha$ trace both the thin star-formation disk and a $\sim 2$ kpc thick disk of extraplanar gas.
Whether the thick disk relates to the extended \ion{H}{1} is
an open question. Low metallicity of the gas at 100 kpc argues against
the most extended gas originating from the midplane, though the metallicity returns
approximately to `normal' within 5 kpc {\bfc of the midplane} \citep{qu19}. It therefore appears that the
$\sim 2$ kpc disk we ascribe to the `chimneys' or `galactic fountain' is distinct
from either the thin star-forming disk and the very extended \ion{H}{1}.

%\subsection{NGC 5907}

The extraplanar distribution of [\ion{C}{2}] for NGC 5907 is similar to that of \ion{H}{1}
(Table~\ref{fittab}).
\citet{allaert15} analyzed archival \ion{H}{1} data and fitted it with a single scale height of 1.2 kpc. 

\subsection{Relationship between [C II] and Star Formation Tracers}

The [\ion{C}{2}] vertical distribution is far wider than the PAH emission (see Fig.~\ref{fig:ngc891pah}). Both [\ion{C}{2}] and PAH emission are
associated with star formation in galaxies \citep{stacey10}.
The PAH emission is likely to be more closely tied to star formation, because the PAH features are
excited by absorption of UV photons that arise from early-type (hence, young) stars. The [\ion{C}{2}]
emission arises from many phases of the ISM including UV-heated star forming regions as well as
diffuse interstellar gas heated by later-type (hence, older) stars. 
If we take the PAH emission as tracing the actual star formation rate (SFR), 
the extraplanar gas has a very different inferred [\ion{C}{2}]/SFR ratio.
In fact, at high $z$ the PAH are excited and the C$^+$ is ionized by star-formation in the disk,
so the local values of `SFR' have relatively little meaning.

The difference in scale heights of [\ion{C}{2}] emission and PAH
indicate that the properties of the ISM and the dominant form of carbon 
changes from the disk to the halo of galaxies. 
The Milky Way average interstellar abundance (outside of molecular clouds)
were derived by comparing the abundance of gas phase C$^+$ measured using {\it Hubble} 
to the interstellar C 
abundance of C/H=$2.4\pm 10^{-4}$ derived from photospheres of B stars, and assuming all C is ionized
but H is neutral along the observed lines of site; this results
in an estimated $58\pm 15$\% of carbon in the gas phase \citep{cardelli96}.
From the brightness of mid-infrared features in the average spectrum of the Milky Way measured by {\it COBE},  \citet{dwek97} found 
$25\pm 5$\% of carbon is locked in the polycyclic aromatic hydrocarbons (PAH). 
Modeling the dust properties, \citet{lidraine} 
also found 25\% of C in grains, with 19\% in small PAH and 6\% in
larger particles modeled as graphite. 

If the [\ion{C}{2}] emission arises from walls of chimneys, then the 
radiation field is likely harder and the dust properties different.
Further, some of the [\ion{C}{2}] emission may arise gas where the 
hydrogen is partially ionized, in the transition between the 
the fully ionized interior and the likely neutral walls. Improved 
images of the extraplanar gas in tracers of neutral gas may allow
to distinguish these components. With the new data presented here,
the [\ion{C}{2}]/\ion{H}{1} ratio increases with distance from the midplane. Compared to the average value within 0.5 kpc of the midplane,
[\ion{C}{2}]/\ion{H}{1} is about 2 times higher at 1 kpc and 3 times higher at 2 kpc.

\section{Conclusions}

Using SOFIA, we observed the 157.7 [\ion{C}{2}] $\mu$m emission line for two nearby, edge-on galaxies to
determine the vertical {\bfc distribution} of the emission with respect to the thin galactic disks.
In both NGC 891 and NGC 5907, the [\ion{C}{2}] distribution contains a thick disk with a scale height of 
$\sim 2$ kpc. The thick disks are intermediate between the very extended \ion{H}{1} 21-cm 
halos of low-metallicity inter/circumgalactic gas and 
the thin disks that contain star-forming regions and molecular gas.
There is some evidence for longitudinal structure in the thick disk.
The extraplanar [\ion{C}{2}] may arise in the walls of `chimneys' that
connect the galaxies' disks to the halo, where material is driven upward
by winds and supernova of massive star forming regions. The walls may be shielded material that has been ejected by the current giant \ion{H}{2} regions or their predecessors.

% SOMETHING ABOUT DARK GAS?

\acknowledgements  
Based in part on observations made with the NASA/DLR Stratospheric Observatory for Infrared Astronomy (SOFIA). SOFIA is jointly operated by the Universities Space Research Association, Inc. (USRA), under NASA contract NNA17BF53C, and the Deutsches SOFIA Institut (DSI) under DLR contract 50 OK 0901 to the University of Stuttgart. 
EXES is supported by NASA agreement NNX13AI85A.
Financial support for the observational and theoretical work in this paper was provided by NASA through award \#06\_0010 issued by USRA.
\facility{SOFIA} 

\bibliographystyle{aasjournal}
\bibliography{wtrbib}

\end{document}